\documentclass{PoS}

\title{Charged Kaon multiplicities of Semi-inclusive DIS off the deuteron target}

\ShortTitle{Kaon multiplicities}

\author{\speaker{Chung Wen Kao}\thanks{This work is supported by MOST Taiwan with the grant no. MOST 105-2112-M-033-004 and MOST 106-2112-M-033-003}\\
        Department of Physics and Centre for High Energy Physics, Chung Yuan Christian University, Chung-Li 32023, Taiwan \\
        E-mail: \email{cwkao@cycu.edu.tw}}

\author{Dong Jing Yang\\
        Department of Physics, National Normal University, Taipei 11677, Taiwan\\
        E-mail: \email{djyang@std.ntnu.edu.tw}}

\author{Wen Chen Chang\\
        Institute of Physics, Academia Sinica, Taipei 11529, Taiwan.\\
        E-mail: \email{changwc@phys.sinica.edu.tw}}

\abstract{We investigate the results of the charged kaon multiplicities off the deuteron target from HERMES and COMPASS experiments. The discrepancy of the two data sets
cannot be explained by different $Q^2$ values. Furthermore we examine the empirical parametrization of the fragmentation functions, DSS2017,
and find that the agreement between the theoretical predictions and the HERMES data is less satisfactory as claimed.}

\FullConference{XXVI International Workshop on Deep-Inelastic Scattering and Related Subjects (DIS2018)\\
		16-20 April 2018\\
		Kobe, Japan}

\begin{document}

\section{Introduction}
Semi-inclusive deep-inelastic scattering, $l+N\to l+h+X$ (SIDIS) plays an unique role in the study of the fragmentation functions (FFs) because
it provides precious information about the flavour dependence of fragmentation functions which cannot be extracted from
$e^{+}e^{-}$ annihilation data. On the other hand, there are attempts to extract the strange-quark Paton Distribution Functions (PDFs)
from the data of the kaon multiplicity of SIDIS off the deuteron target but not without controversy~\cite{strange1,strange2,strange3}, based on the recent data
provide by HERMES collaboration at DESY ~\cite{Airapetian:2012ki,Airapetian:2013zaw}.
The leading order (LO) formula of the kaon multiplicity off the deuteron target is given as,
\begin{eqnarray}
M^{K}_{D}(x,Q^2)&\equiv&M^{K^{+}}_{D}(x,Q^2)+M^{K^{-}}_{D}(x,Q^2)=\frac{dN^{K}(x,Q^2)}{dN^\mathrm{DIS}(x,Q^2)}
\cr
&=&\frac{\sum_{q}e_{q}^2\left[q^{p}(x,Q^2)+\bar{q}^{p}(x,Q^2)+q^{n}(x,Q^2)+\bar{q}^{n}(x,Q^2)\right]\int^{z_\mathrm{max}}_{z_\mathrm{min}}D_{q}^{K}(z,Q^2)dz}
{\sum_{q}e_{q}^2\left[q^{p}(x,Q^2)+\bar{q}^{p}(x,Q^2)+q^{n}(x,Q^2)+\bar{q}^{n}(x,Q^2)\right]}.
\label{Eq:ori2}
\end{eqnarray}
Here $q=(u,d,s)$ and $e_q$ are the quark flavours and the corresponding electric charges, respectively.
Besides $q^{i}(x,Q^2)$ with $i \in \{p,n\}$ are the relevant nucleon PDFs with momentum fraction $x$ and
momentum transfer squared $Q^2$. Notice the superscripts $p$ and $n$ denote proton and neutron.
The $z$ is the momentum fraction of the initial quark in the fragmented hadron and $z_\mathrm{max}$ and $z_\mathrm{min}$ are usually set by the
experimental acceptance. Finally $D_{q}^{K}$ in Eq.~(\ref{Eq:ori2}) is defined in terms of FFs as well and takes the the
following form $D_{q}^{K}(z,Q^2)=D_{q}^{K^{+}}(z,Q^2)+D_{q}^{K^{-}}(z,Q^2)$. If the isospin symmetry is assumed then it is possible to obtain the strange PDFs
from Eq.~(\ref{Eq:ori2}). Actually in our previous work we find that such an extraction crucially depends on the choice of the fragmentation functions ~\cite{Yang:2015avi}.
Furthermore we also point out that such an extraction actually can be carried out by the pion multiplicities data and there is serious tension between
the results from the pion multiplicity and kaon multiplicity~\cite{Yang:2015avi}.
Nevertheless all these studies are based on the leading order (LO) formula.
Hence it is necessary to investigate the hadron multiplicities according to
the next-leading-order (NLO) formula. The NLO formula of SIDIS is as follows,

\begin{eqnarray}
\sigma^{h}(x,z)&=&\sum_{f}e_f^2 q_{f}\otimes D^{h}_{q_f}
+\frac{\alpha_s}{2\pi}\left(e_f^2 q_{f}\otimes {\cal C}_{qq}\otimes D^{h}_{q_f}
+e_f^2 q_{f}\otimes {\cal C}_{qg}\otimes D^{h}_{G}
+\frac{\alpha_s}{2\pi}G\otimes {\cal C}_{gq}\otimes \sum_{q_f}e_f^2D^{h}_{q_f}\right).
\cr
&& q\otimes{\cal C}\otimes D(x,z)\equiv \int_{x}^{1}\frac{dx'}{x'}\int_{z}^{1}\frac{dz'}{z'}q\left(\frac{x}{x'}\right){\cal C}(x',z')D\left(\frac{z}{z'}\right).
\label{Eq:NLO}
\end{eqnarray}

\begin{figure}
\begin{tabular}{ccccccc}
\includegraphics[width=.13\textwidth]{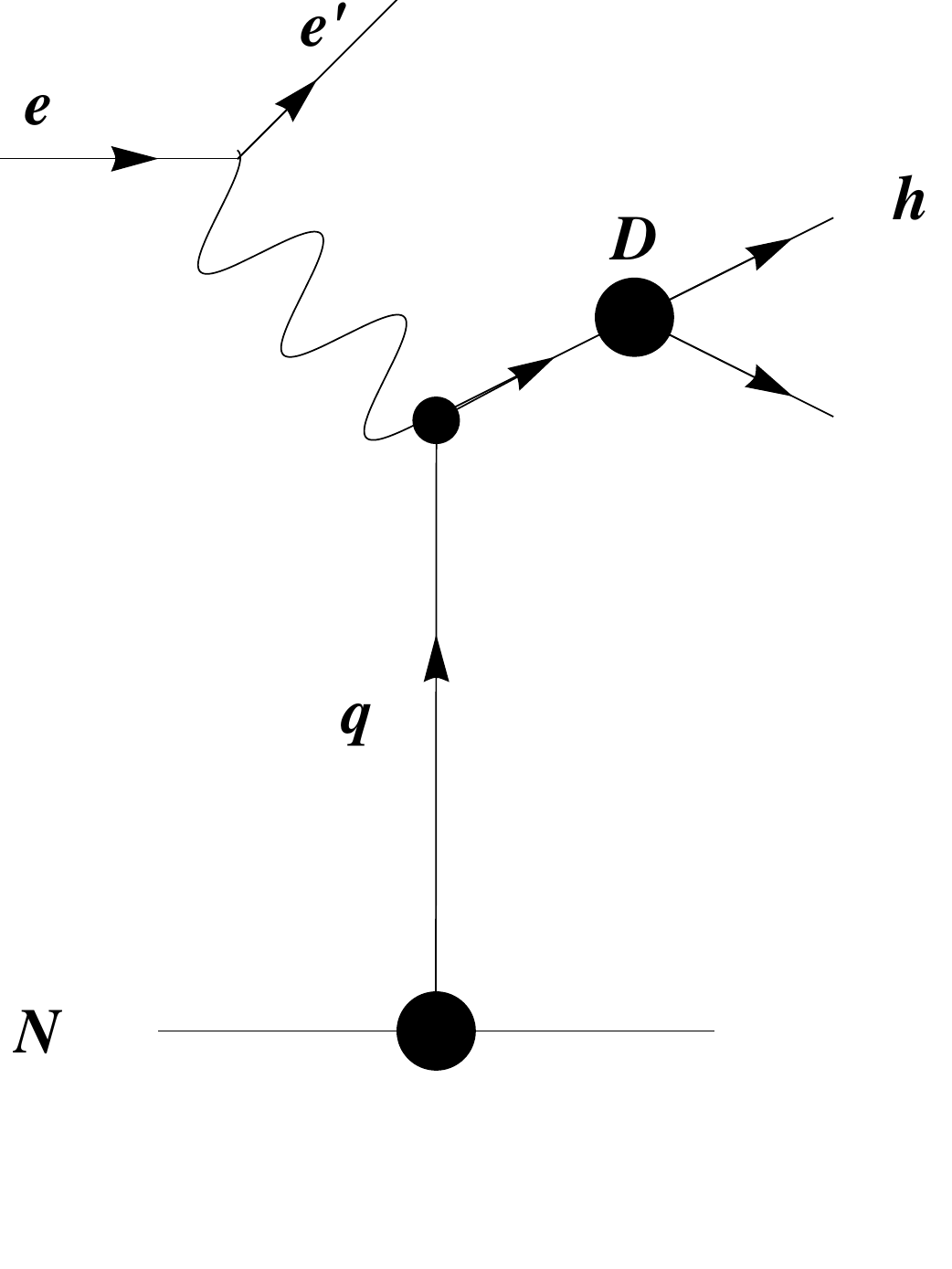}
\includegraphics[width=.13\textwidth]{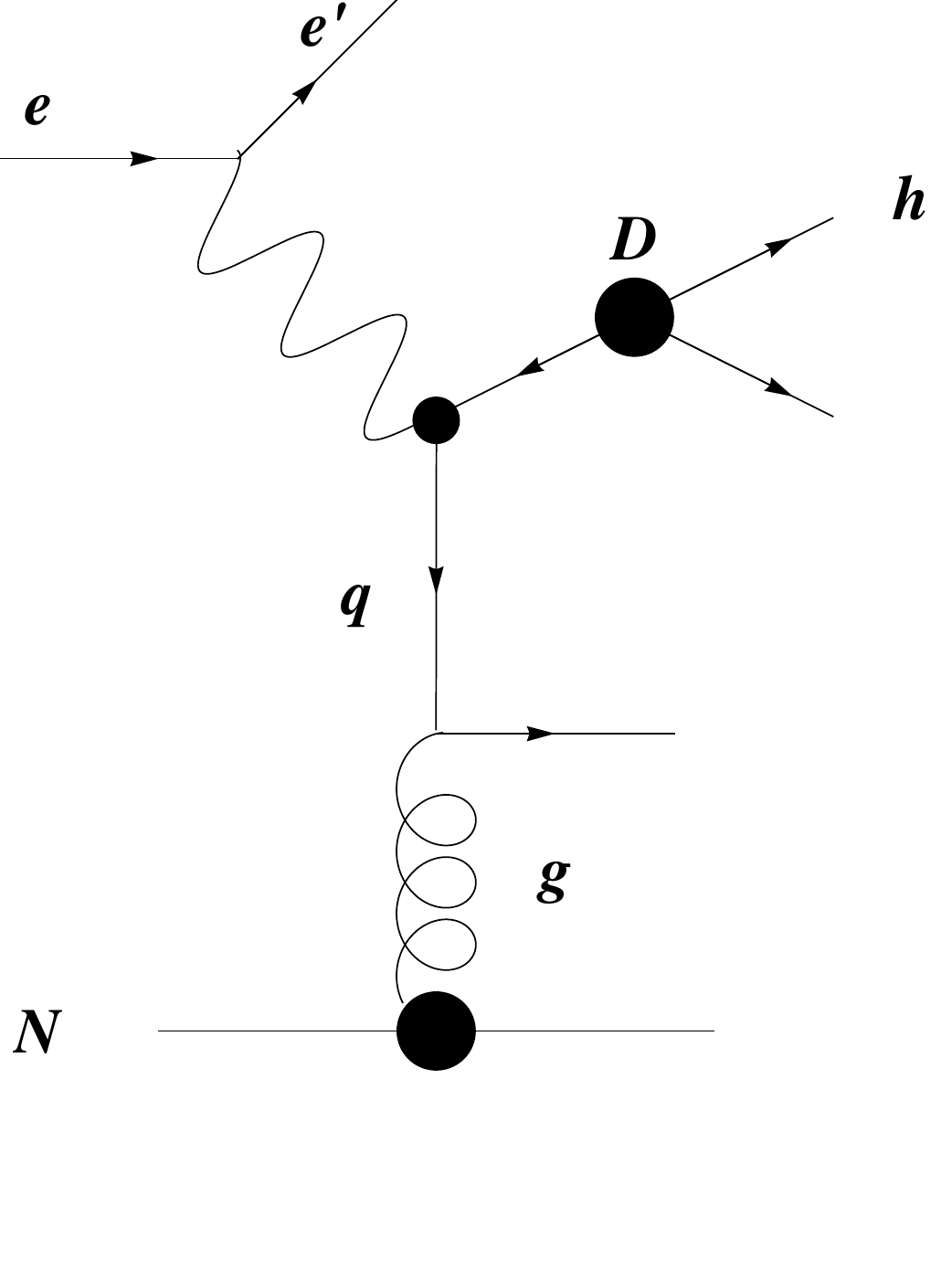}
\includegraphics[width=.13\textwidth]{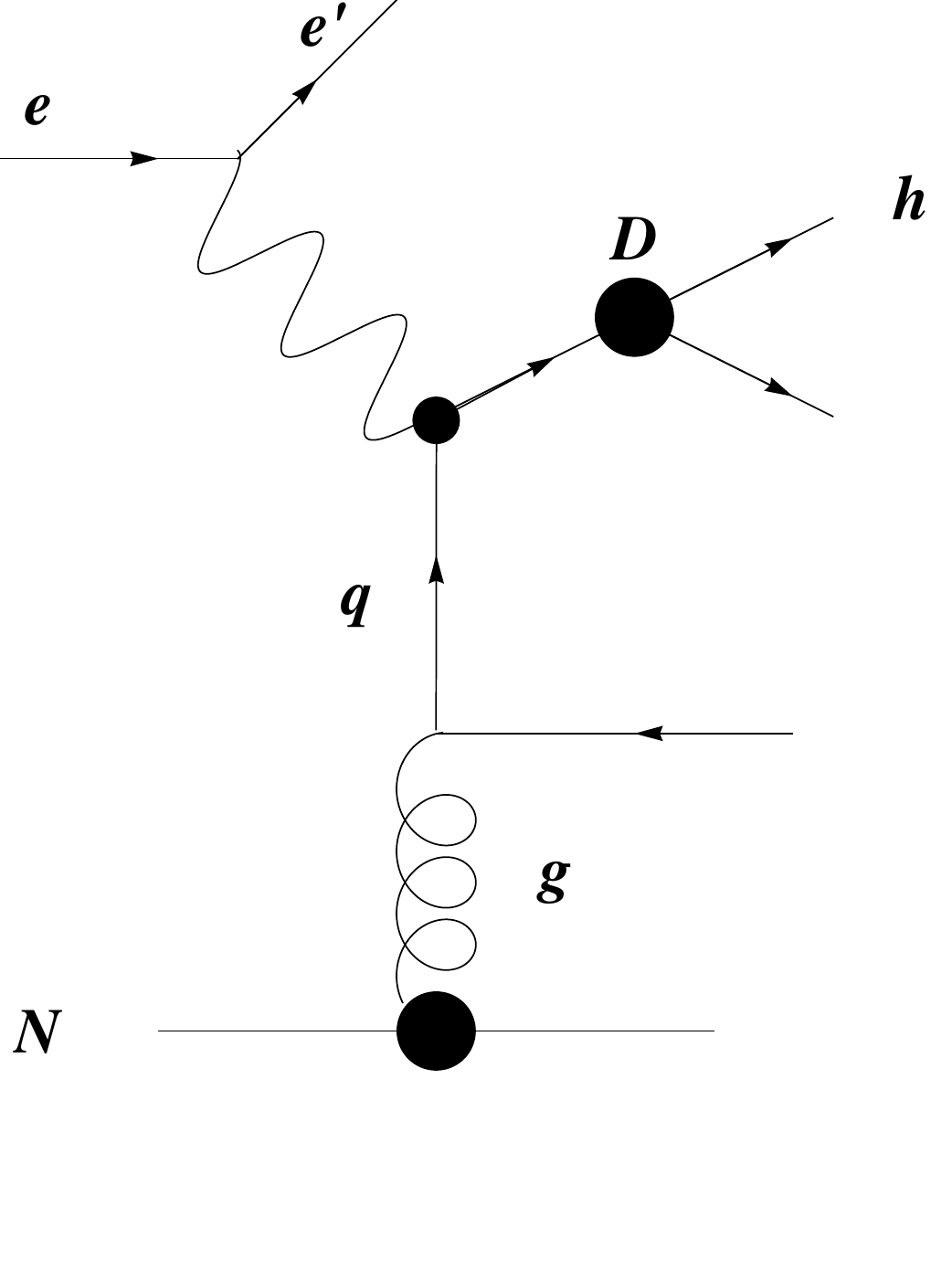}
\includegraphics[width=.13\textwidth]{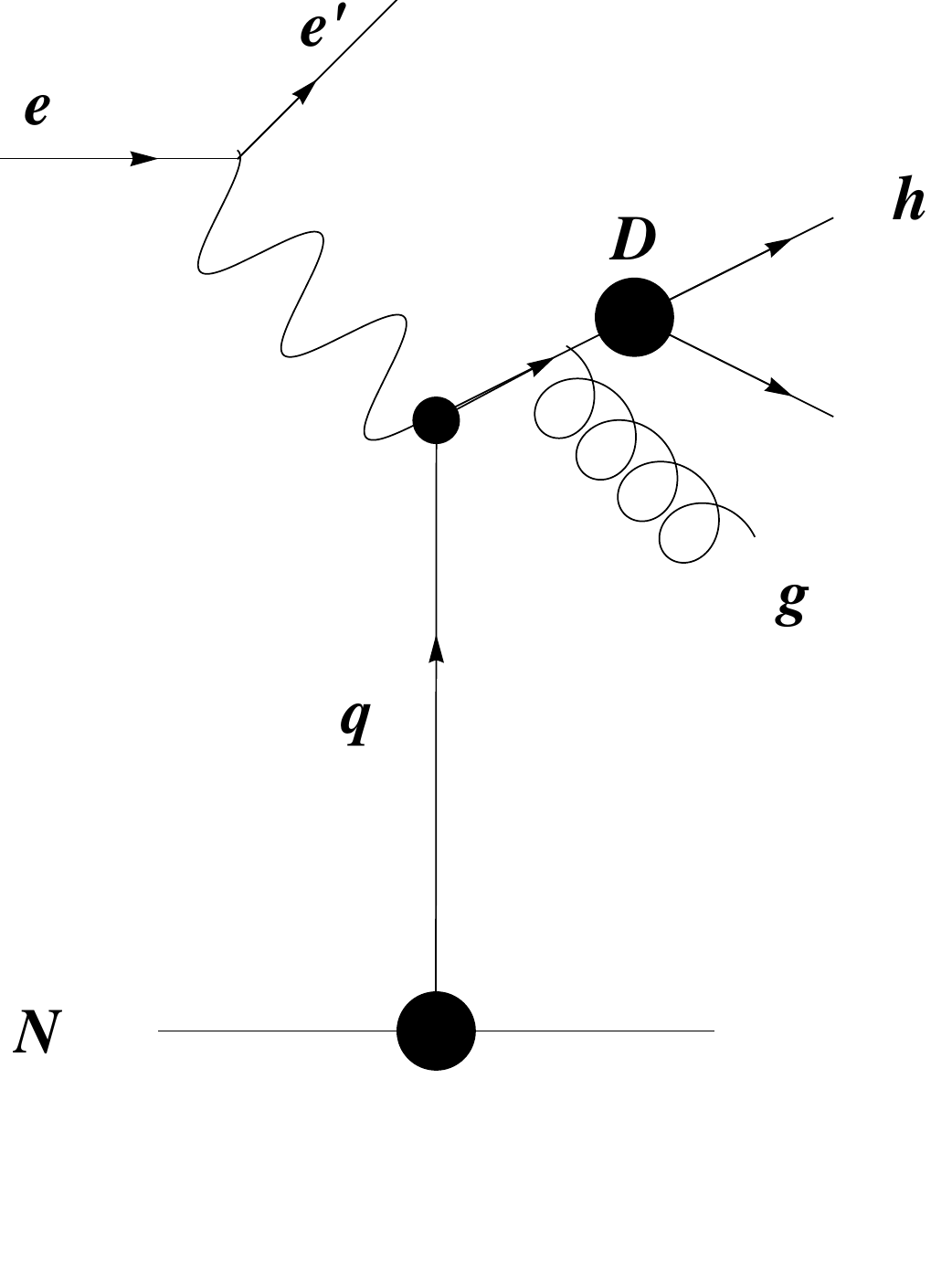}
\includegraphics[width=.13\textwidth]{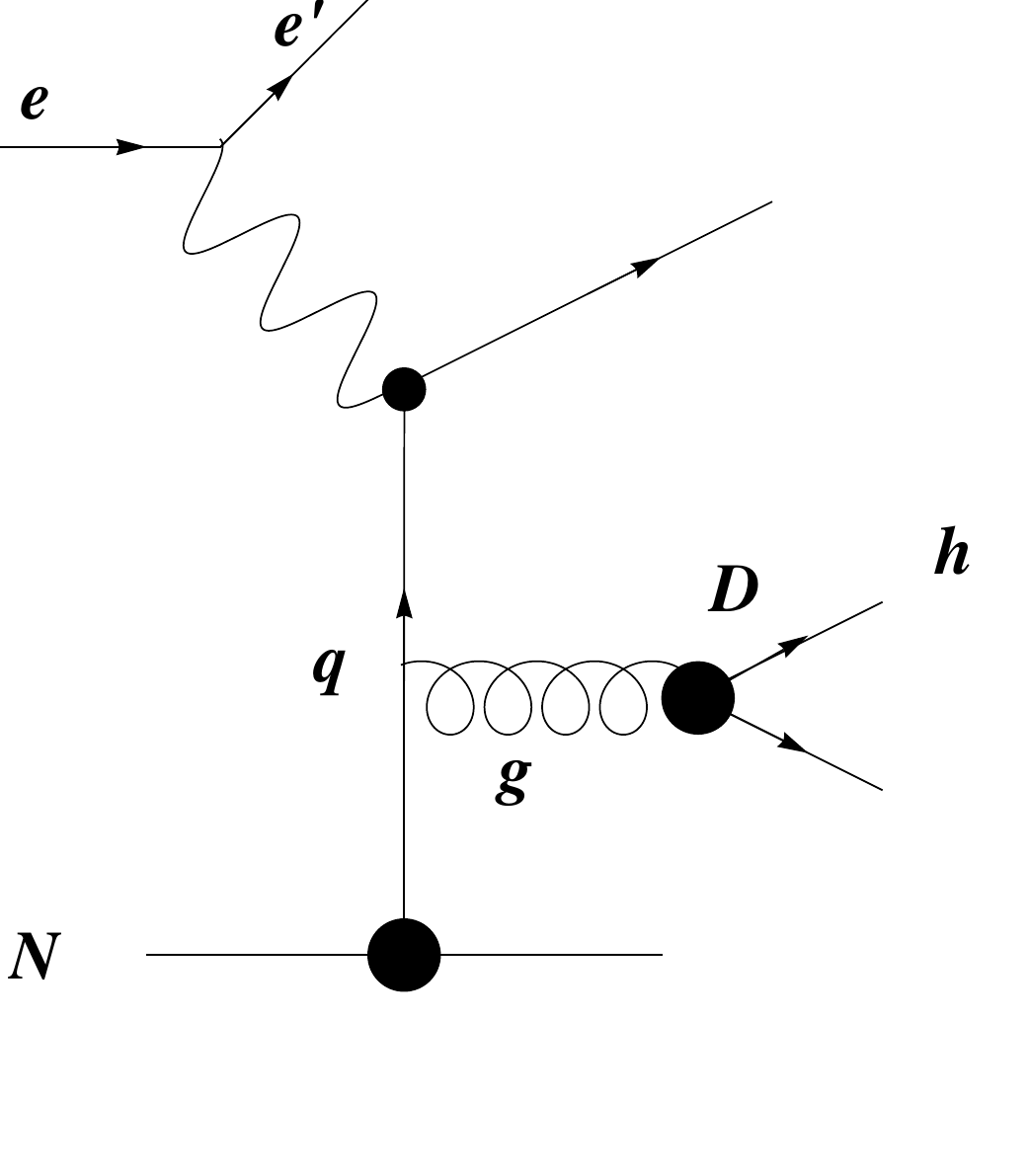}
\includegraphics[width=.13\textwidth]{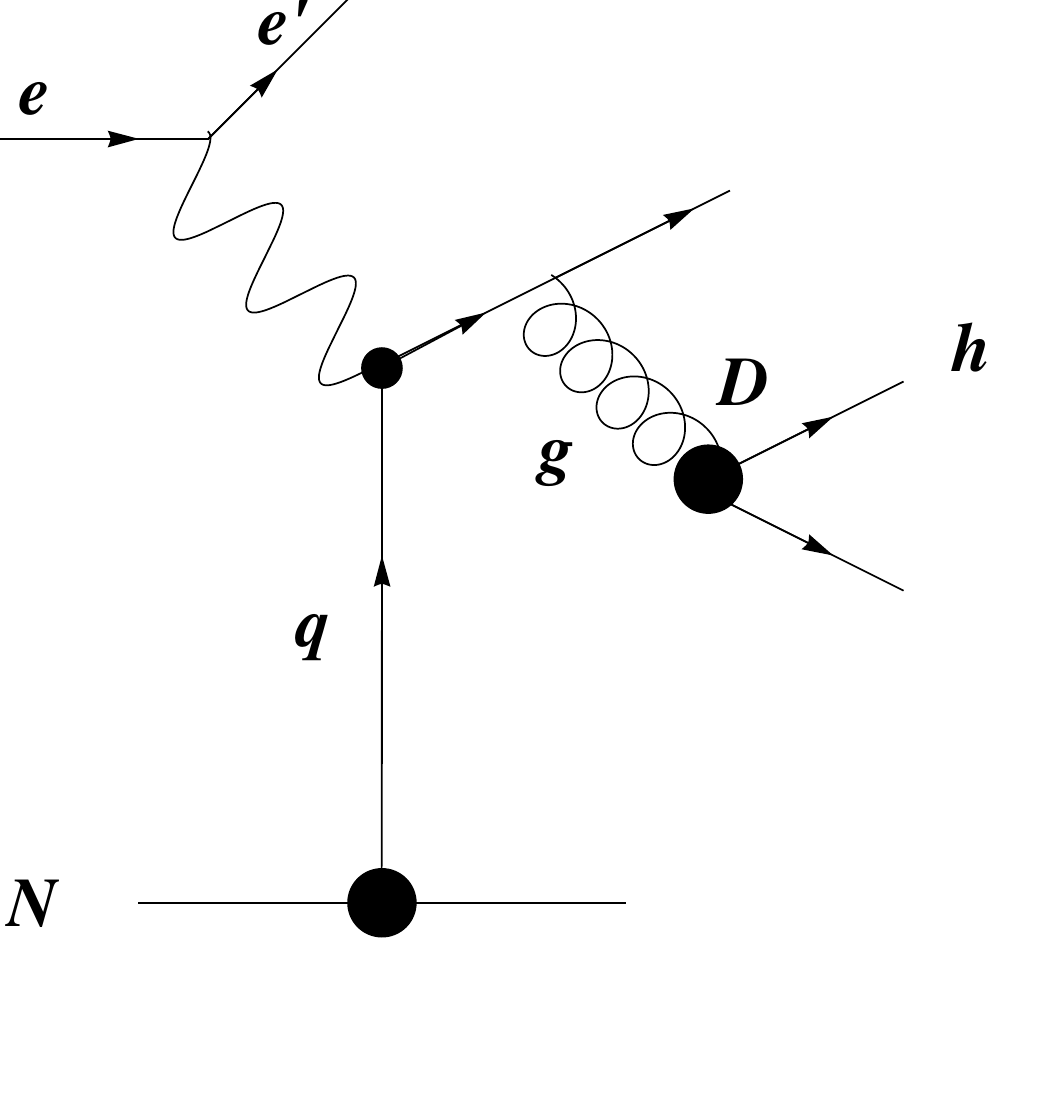}
\includegraphics[width=.13\textwidth]{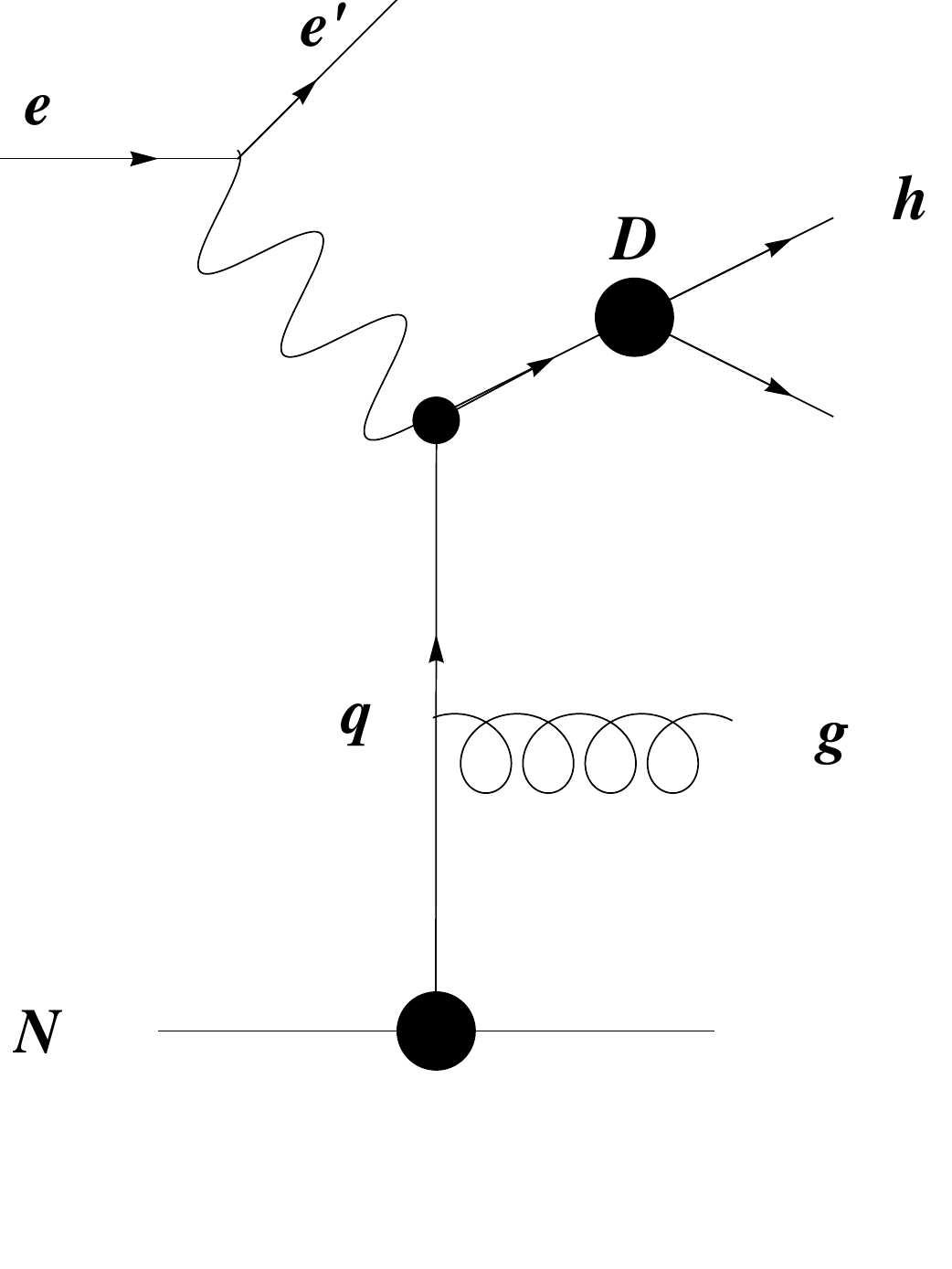}
\end{tabular}
\caption{This is the set of the Feynman diagrams of SIDIS up to NLO. The LO diagram is the leftmost one. The others are the NLO diagrams.}
\label{NLO}
\end{figure}

The associated Feynman diagrams are depicted in Fig.(~\ref{NLO}). ${\cal C}$ are the splitting functions~\cite{Furmanski:1981cw}.
In principle, one would still extract the strange quark distributions from the kaon multiplicity data with the above formula if accurate fragmentation functions are available.
However so far the fragmentation functions are still far away from perfect and the strange PDFs are mainly determined by other processes. Instead to extract the strange PDFs
from the HERMES data of the kaon multiplicities, it is more natural to make global fits of the fragmentation functions from the data
of SIDIS by including the HERMES data ~\cite{Airapetian:2012ki,Airapetian:2013zaw}.

\section{Tension between the results of the Kaon multiplicities of HERMES and COMPASS Data}
\begin{figure}
\begin{tabular}{cc}
\includegraphics[width=.5\textwidth]{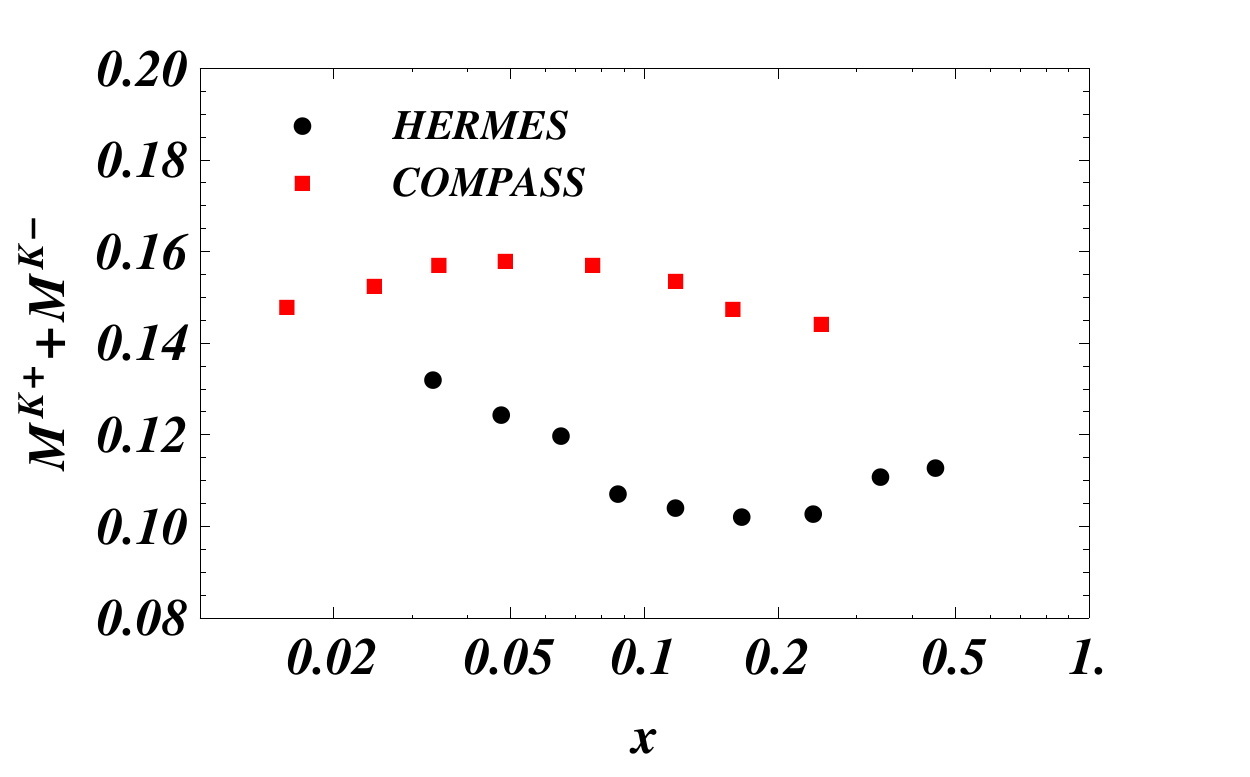}
\includegraphics[width=.5\textwidth]{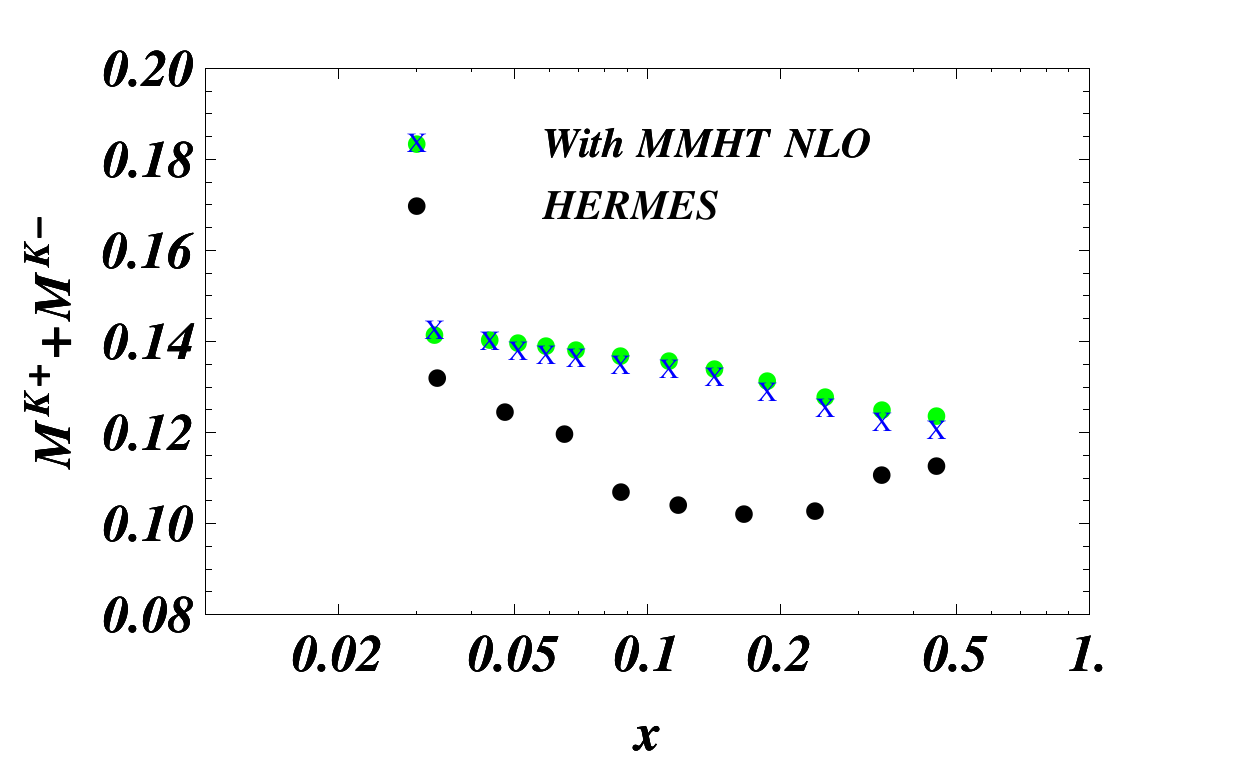}
\end{tabular}
\caption{(a)Experimental results of charged kaon multiplicities from HERMES ~\cite{Airapetian:2012ki,Airapetian:2013zaw} and COMPASS ~\cite{Adolph:2016bwc} experiments.
(left panel). (b) Theoretical predictions with the DSS2017 parametrization and MMHT PDFs at HERMES kinematics compared with the experimental data.}
\label{data}
\end{figure}

Recently the COMPASS results of the charged kaon multiplicities of SIDIS off the deuteron target have been published ~\cite{Adolph:2016bwc}.
From Fig.(\ref{data})a the COMPASS results obviously deviate from the HERMES ones. However such difference may be
just superficial because the corresponding $Q^2$ of two sets of data are not the same. Actually each data point has its own
$Q^2$ value. One can find those kinematic conditions in Table.~(\ref{Table}). To check the consistency between the two data sets, one needs to
know whether the effect of QCD evolution is able to account for these differences.
However it is difficult, if not impossible, to apply QCD evolution
on the experimental data directly. Instead, it is more practical to choose reliable FFs and PDFs and apply the NLO formula
, Eq.~(\ref{Eq:NLO}) to obtain the theoretical predictions at the precise $Q^2$ scales.

\begin{figure}
\begin{tabular}{cc}
\includegraphics[width=.5\textwidth]{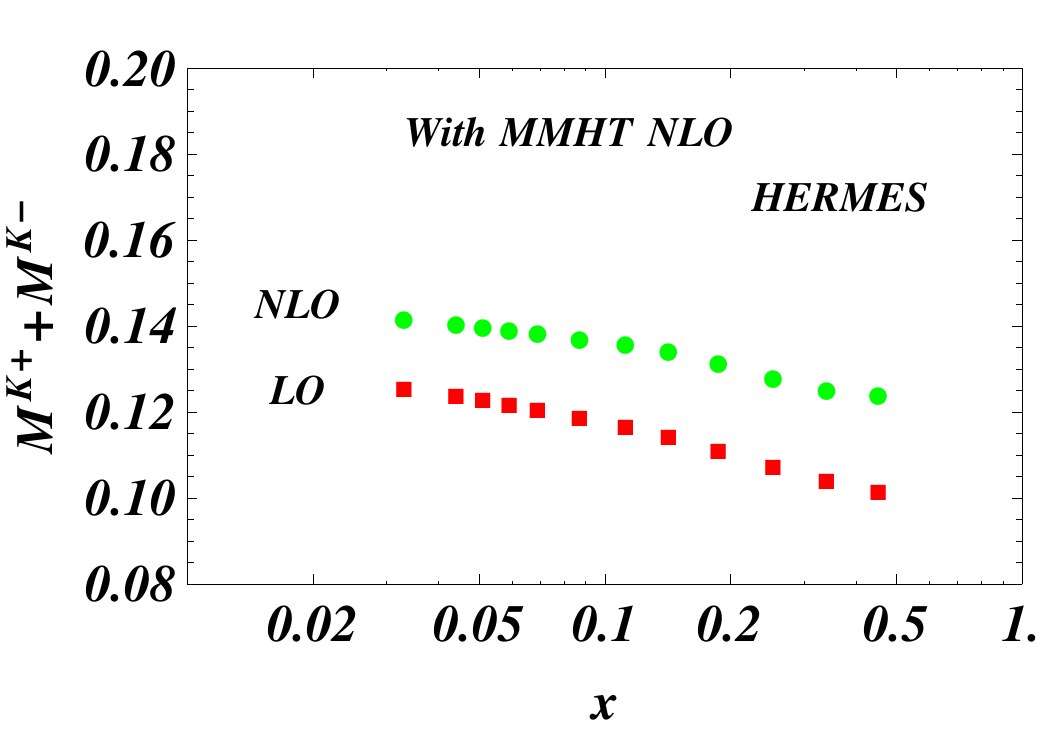}
\includegraphics[width=.5\textwidth]{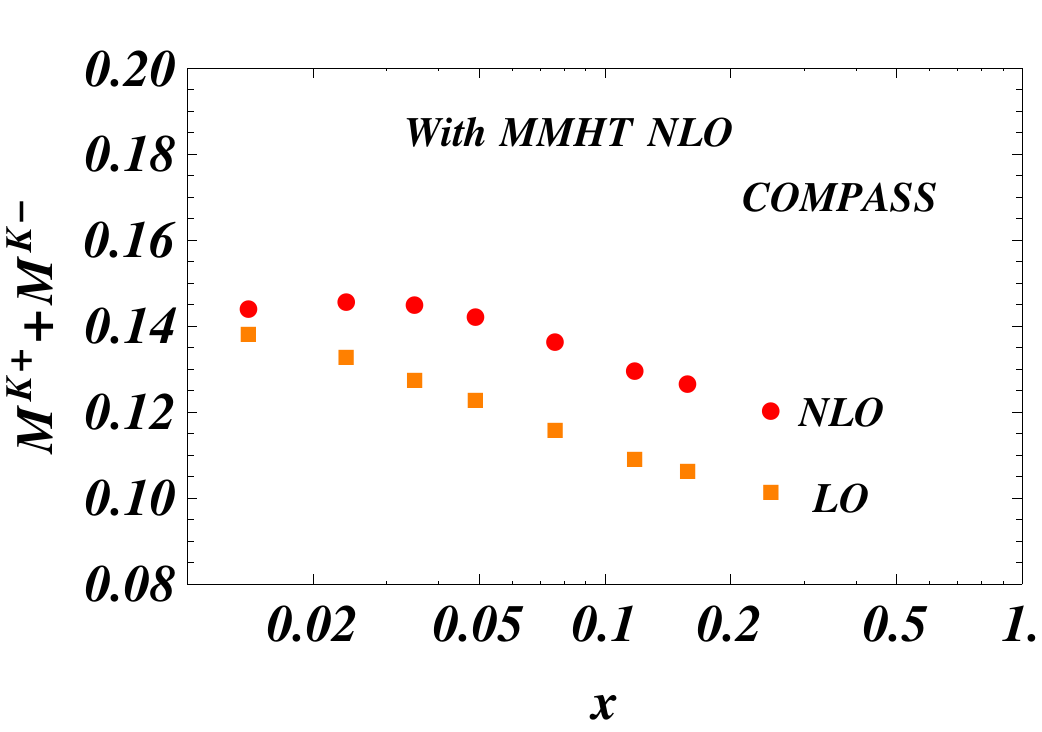}
\end{tabular}
\caption{The LO and NLO theoretical calculations of the kaon multiplicities at the kinematic conditions of HERMES (left panel)
and COMPASS experiments (right panel).}
\label{DSS2017}
\end{figure}

\begin{figure}
\begin{tabular}{cc}
\includegraphics[width=.5\textwidth]{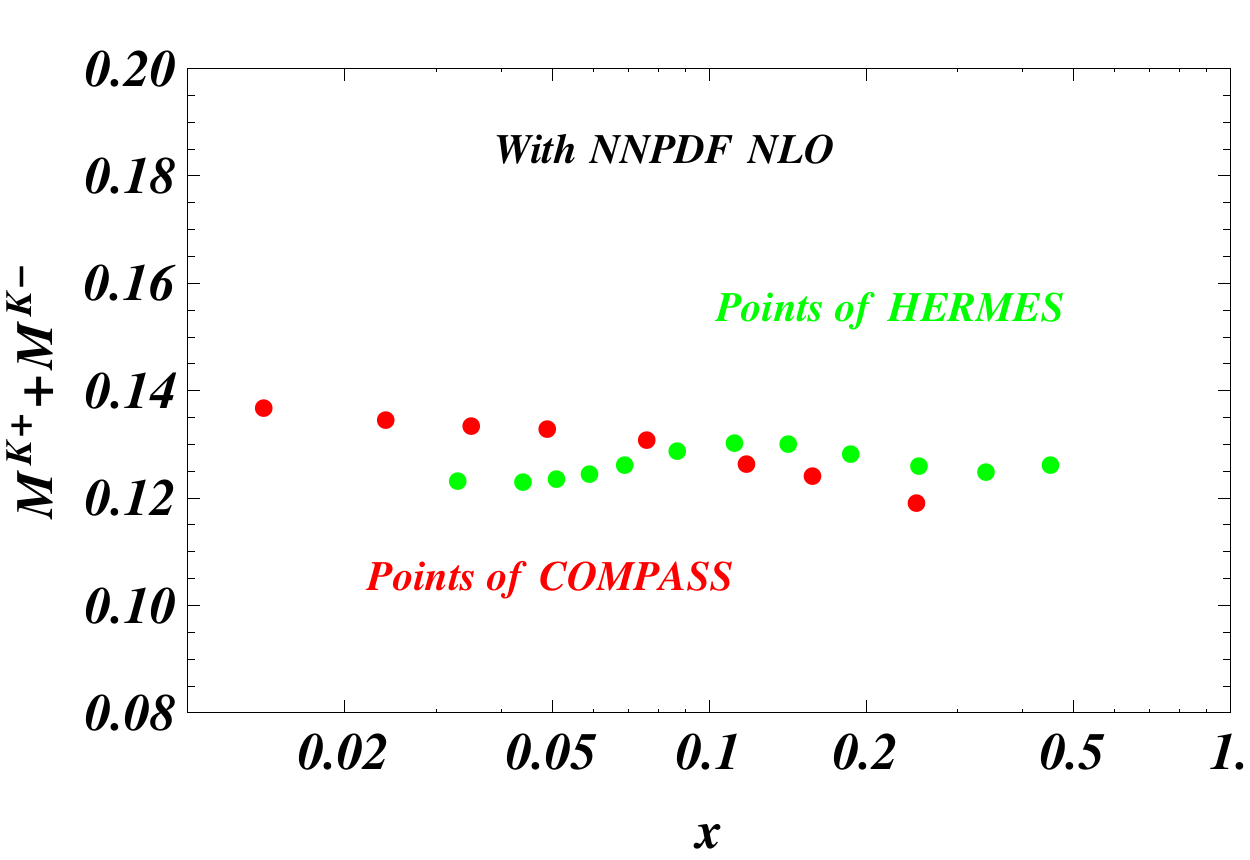}
\includegraphics[width=.5\textwidth]{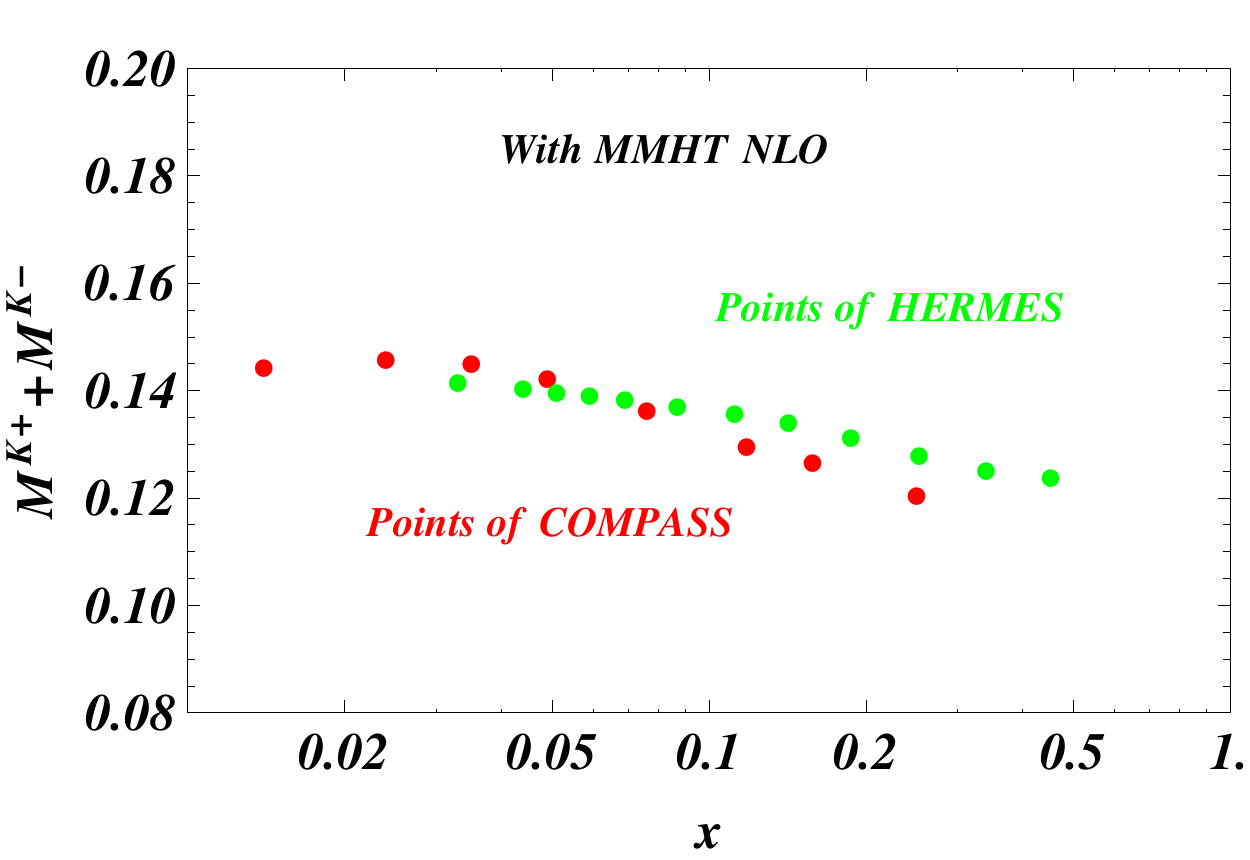}
\end{tabular}
\caption{Theoretical predictions of the kaon multiplicities at the kinematic conditions of HERMES and COMPASS experiments with different PDFs.
NNPDF3.0 (left) and HHMT2014 (right).}
\label{PDF}
\end{figure}

Last year DSS2017 parametrization became available ~\cite{deFlorian:2017lwf}. This parametrization is the updated version of the previous one~\cite{deFlorian:2007ekg}. The main difference between the new one and the old one is the updating of inclusion of the COMPASS data~\cite{Adolph:2016bwc}. Their differential kaon multiplicity predictions are compared with the COMPASS and HERMES data. It is claimed their parametrization is able to describe both of the data sets well simultaneously ~\cite{deFlorian:2007ekg}. Therefore we use DSS2017 parametrization with MMHT PDFs~\cite{Harland-Lang:2014zoa} (as used in~\cite{deFlorian:2007ekg}),
to compare with the results in Fig.~(\ref{data}). Our results are presented in Fig.~(\ref{DSS2017}). We demonstrate the LO and NLO result at the HERMES and COMPASS kinematics, respectively. It is obvious that the difference between LO and NLO results are significant and their
agreement with the experimental data is far from satisfactory.
\begin{table}
\begin{tabular}{|c|c|c||c|c|c||c|c|c|}
\hline
HERMES&$x$&$Q^2(GeV^2)$&HERMES&$x$&$Q^2(GeV^2)$&COMPASS&$x$&$Q^2(GeV^2)$\\
\hline
$A$&0.0033&1.1931&$J$&0.253&5.19&$A$&0.0085&1.1709\\
\hline
$B$&0.044&1.3822&$K$&0.34&7.4768&$B$&0.0157&1.454\\
\hline
$C$&0.051&1.42&$L$&0.452&10.2355&$C$&0.0247&2.1489\\
\hline
$D$&0.059&1.50&&&&$D$&0.0345&3.0170\\
\hline
$E$&0.069&1.59&&&&$E$&0.0487&4.2476\\
\hline
$F$&0.087&1.7278&&&&$F$&0.0765&6.6756\\
\hline
$G$&0.112&2.05&&&&$G$&0.1176&10.2629\\
\hline
$H$&0.142&2.67&&&&$H$&0.1581&11.8938\\
\hline
$I$&0.187&3.63&&&&$I$&0.2502&20.0857\\
\hline
\end{tabular}
\caption{Kinematics conditions of HERMES and COMPASS experiments.}
\label{Table}
\end{table}
We also apply another PDFs, NNPDF ~\cite{NNPDF} to do the similar calculation.
The results are shown in Fig.~(\ref{PDF}). It is clear to see that the difference between the theoretical predictions at the HERMES and COMPASS kinematics are much closer to each other than the experimental data in Fig.~(\ref{data})a, in both cases of MMHT and NNPDF.
For further investigation, one needs to clarify the issue that whether DSS2017 parametrization can describe the COMPASS as well as HERMES data
in term of the differential charged kaon multiplicities~\cite{deFlorian:2007ekg}.
Since DSS2017 paper has demonstrated their result agree with the experimental data well before integrating $z$ values,
it is necessary to examine the procedure of integrating the kaon multiplicities in HERMES and COMPASS results.
Hence in the next section we will take a closer look at the DSS2017 parametrization.

\section{Take a closer look at the DSS2017 parametrization of the fragmentation functions}

\begin{figure}
\begin{tabular}{cc}
\includegraphics[width=.5\textwidth]{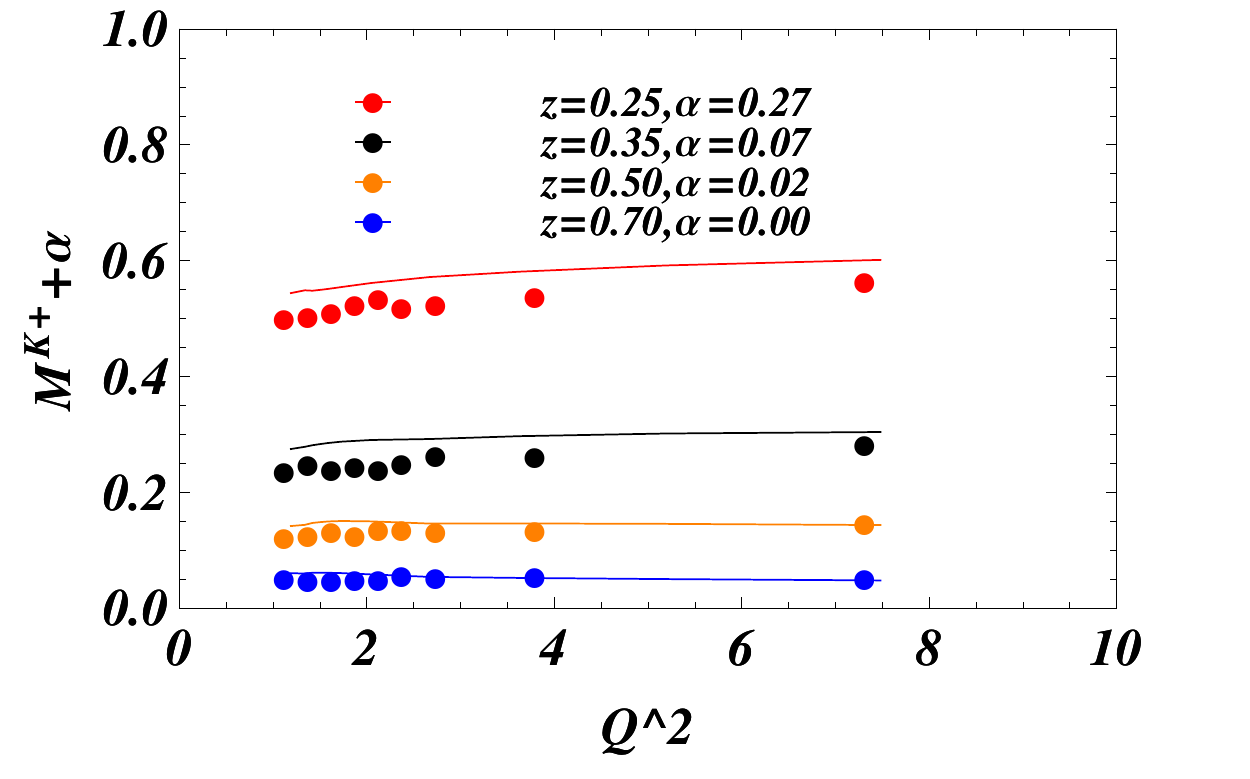}
\includegraphics[width=.5\textwidth]{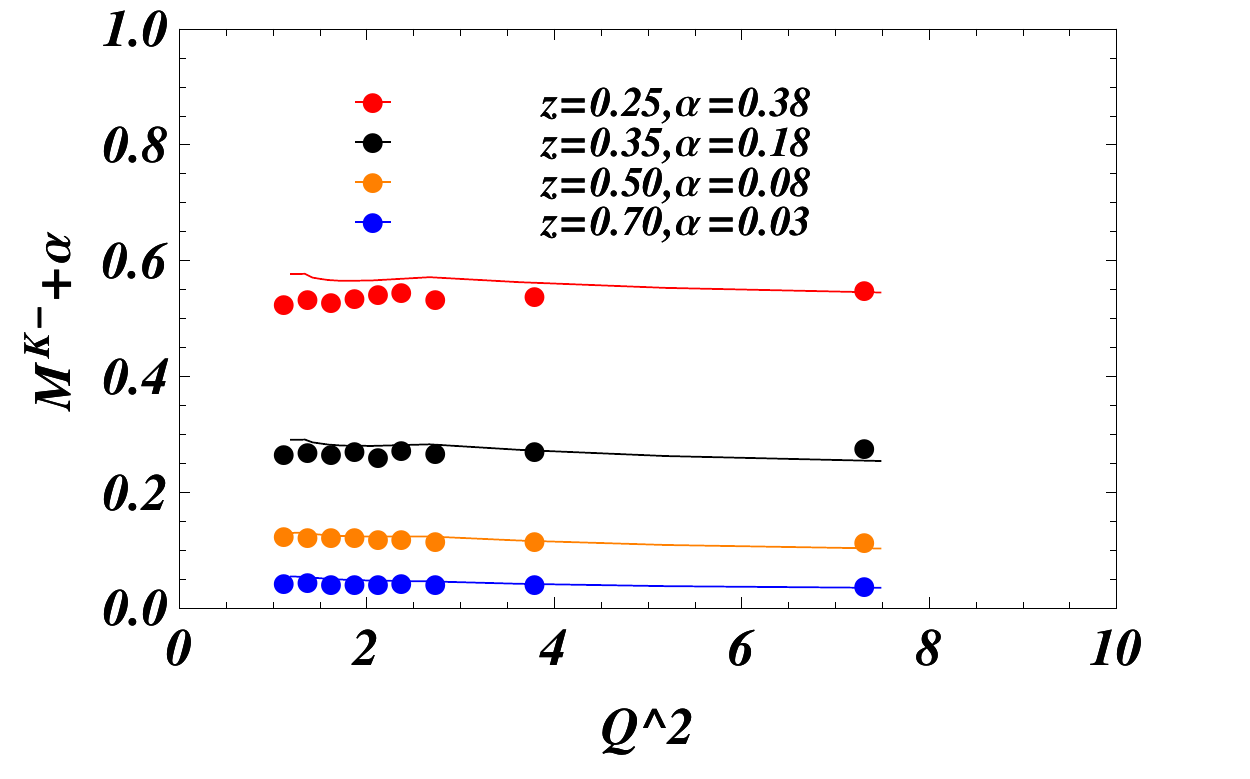}
\end{tabular}
\caption{The kaon multiplicities as functions of $Q^2$ at different $z$ ranges.}
\label{line}
\end{figure}

In Ref~\cite{deFlorian:2007ekg}, It has shown that the agreement between the DSS2017 differential result and the HERMES experimental data seems to be excellent.
Unfortunately it is mere illusory. When one changes the logarithmic scale into the linear one, the seemingly excellent agreement is gone.
One can easily figure it out by observing Fig.~(\ref{line}). The offset $\alpha$ is just added to make comparison easier to be observed.
The curves in ~Fig.~(\ref{line}) are just the lines connecting the theoretical result at each data point.
In the $K^{+}$ case, one find the theoretical curves are larger than the experimental data about $10\%$ when $z$ is smaller than $0.35$.
This trend is held for all the data points even they are taken at different $x$ and $Q^2$ values.
In the $K^{-}$ case, the differences between the curve and the data become smaller and at the large $z$ regime the data is even larger than the curve.
Summing over $z$ range one obtains the results presented in Fig.~(\ref{sum}). The crosses stand on the result obtained by summing over the $z$ bins and the circle
point stands on the theoretical predictions obtained by integrating over $z$ directly. (These notations are also used in Fig.~({\ref{data})b.)
It is not surprising that theoretical predictions are larger than the HERMES data in Fig.~(\ref{sum}).
The difference for the $K^{+}$ case becomes larger when $Q^2$ increases. On the other hand, the difference for the $K^{-}$ case
is smaller than the $K^{+}$ case and at the largest $Q^2$ point our result becomes smaller than the HERMES data point.
When we sum over the $K^{+}$ and $K^{-}$ cases, the result is depicted in Fig.~(\ref{data})b.
\begin{figure}
\begin{tabular}{cc}
\includegraphics[width=.5\textwidth]{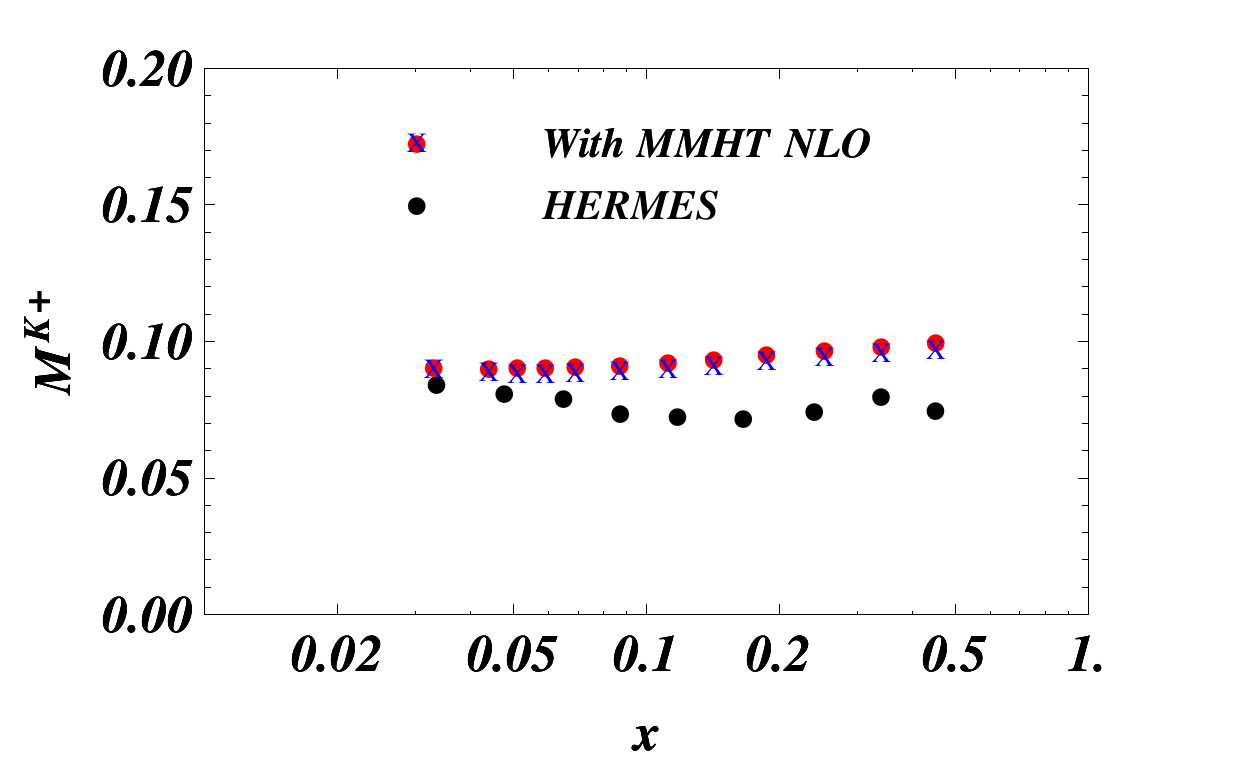}
\includegraphics[width=.5\textwidth]{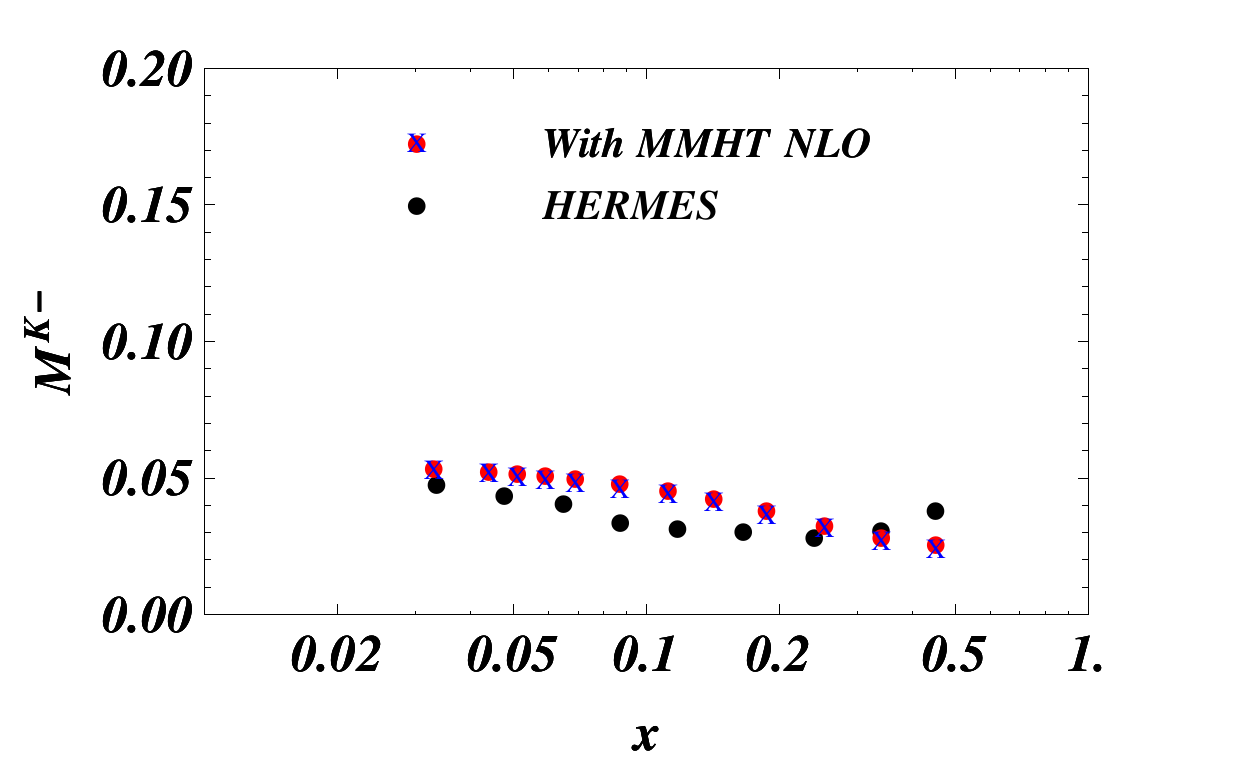}
\end{tabular}
\label{sum}
\caption{Comparison between the theoretical and experimental results of the kaon multiplicities at the kinematics of HERMES data. The left panel is for
$K^{+}$ case and the right panel is for $K^{-}$ case.}
\end{figure}
It turns out that the difference between the theoretical result and the HERMES data is most significant in the middle-$x$ region with $Q^2$ located between
$2-4$ GeV$^2$. The difference would reach $40\%$! In general the HERMES results of the kaon multiplicity are smaller than what one expects
if DSS2017 parametrization is used. Hence the claim that the DSS2017 parametrization is able to describe both of the data sets well simultaneously in~\cite{deFlorian:2007ekg}
is not confirmed in our study.
\section{Conclusion and Outlook}
 We investigate the charged kaon multiplicities off the deuteron target from HERMES and COMPASS experiments. The discrepancy of the two data
 cannot be explained by the different $Q^2$ values.
 Furthermore we find that the agreement between the NLO theoretical predictions with the DSS2017 parametrization
 and the HERMES data are less satisfactory as claimed.
 We plan to study the similar issues for the COMPASS data and extend our analysis to other hardon multiplicities of the SIDIS off the deuteron and proton target and hope our study will shed some light on the cause of the discrepancy of the HERMES and COMPASS data.

\end{document}